\documentclass{sf2a-conf2010}
\usepackage{graphicx}
\usepackage{hyperref}
\usepackage[]{natbib}

\begin{document}
\TitreGlobal{SF2A 2010}
%
%%-----------------------------
%%      the top matter
%%-----------------------------
\title{Chemically peculiar A/F stars in open clusters of the Milky Way}
\author{Gebran, M.}\address{Departament d'Astronomia i Meteorologia, Universitat de Barcelona, c/ Mart\`i i Franqu\`es, 1, 08028 Barcelona, Spain.\\ Present affiliation:Department of Physics and Astronomy, Notre Dame University-Louaize, PO Box 72, Zouk Mikael, Lebanon.}
\author{Monier, R.}\address{Laboratoire Universitaire d'Astrophysique de Nice, UMR 6525, Universit\'e de Nice - Sophia Antipolis, Parc Valrose, 06108 Nice Cedex 2, France.}
\runningtitle{A/F stars in open clusters}
\setcounter{page}{237}
% Keep this line, even if the page will be settled afterwards..

\index{Gebran, M.}
\index{Monier, R.}
% Repeat the authors here, this will help to make the final index

\maketitle
\begin{abstract}
%
% Warning!  within the abstract:
% - do not use macros. 
% - do not use commands like: \cite, \citet, \citep ... etc.
Abundance anomalies have been determined at the surface of many field and open cluster A and F dwarfs. These abundance anomalies are most likely caused by microscopic diffusion at work within the stable envelopes of A stars. However diffusion can be counteracted by several other mixing processes such as convection, rotational mixing and mass loss. We present a short review of the surface abundance patterns of A/F stars in the Pleiades (100 Myr), Coma Berenices (450 Myr) and Hyades (650 Myr) open clusters. Real star-to-star variations of the abundances were found for several chemical elements in the A dwarfs in these clusters. The derived abundances are then compared to evolutionary models from the Montreal group.  These comparisons strongly suggest the occurence of hydrodynamical processes at play within the
radiative zones of these stars and hindering the effects of microscopic diffusion (mixing processes/mass loss). In the frame of Gaia mission, simulations are presented that predict the number of A stars and open clusters that Gaia will observe in the Galaxy. 
\end{abstract}
% Insert the keywords (to appear in the ADS indexing)
\begin{keywords}
stars: abundances - stars: main sequence - stars: rotation - diffusion - Galaxy: open clusters and associations
\end{keywords}
\section{Introduction}
%---------------------
Abundance studies of A stars have mostly focused on the Chemically Peculiar A stars as their low projected rotationnal velocities facilitate abundance determinations. In contrast, little is known about the chemical composition of normal A stars. Star to star abundance variations have been found for a handful of normal A field stars (Lemke 1990, Hill \& Landstreet 1993 and Hill 1995), definitely showing that these normal stars do not have a solar chemical composition.\\
Stars in open clusters originate from the same interstellar material (i.e. they have the same age and the same initial chemical composition) and as such are very useful to test the predictions of evolutionary models. Varenne \& Monier (1999), Gebran et al. (2008), Gebran \& Monier (2008) and Gebran et al. (2010) 
also found significant star to star abundance variations for most A stars members of the Pleiades, Coma Berenices and the Hyades. Elemental radiative diffusion is the main process to account for anomalous abundances in Am stars. However turbulence, mass loss, accretion and meridional circulation may play a role as well.  

\section{Results}
\begin{figure}[h]
   \centering
 
 \includegraphics[scale=0.42]{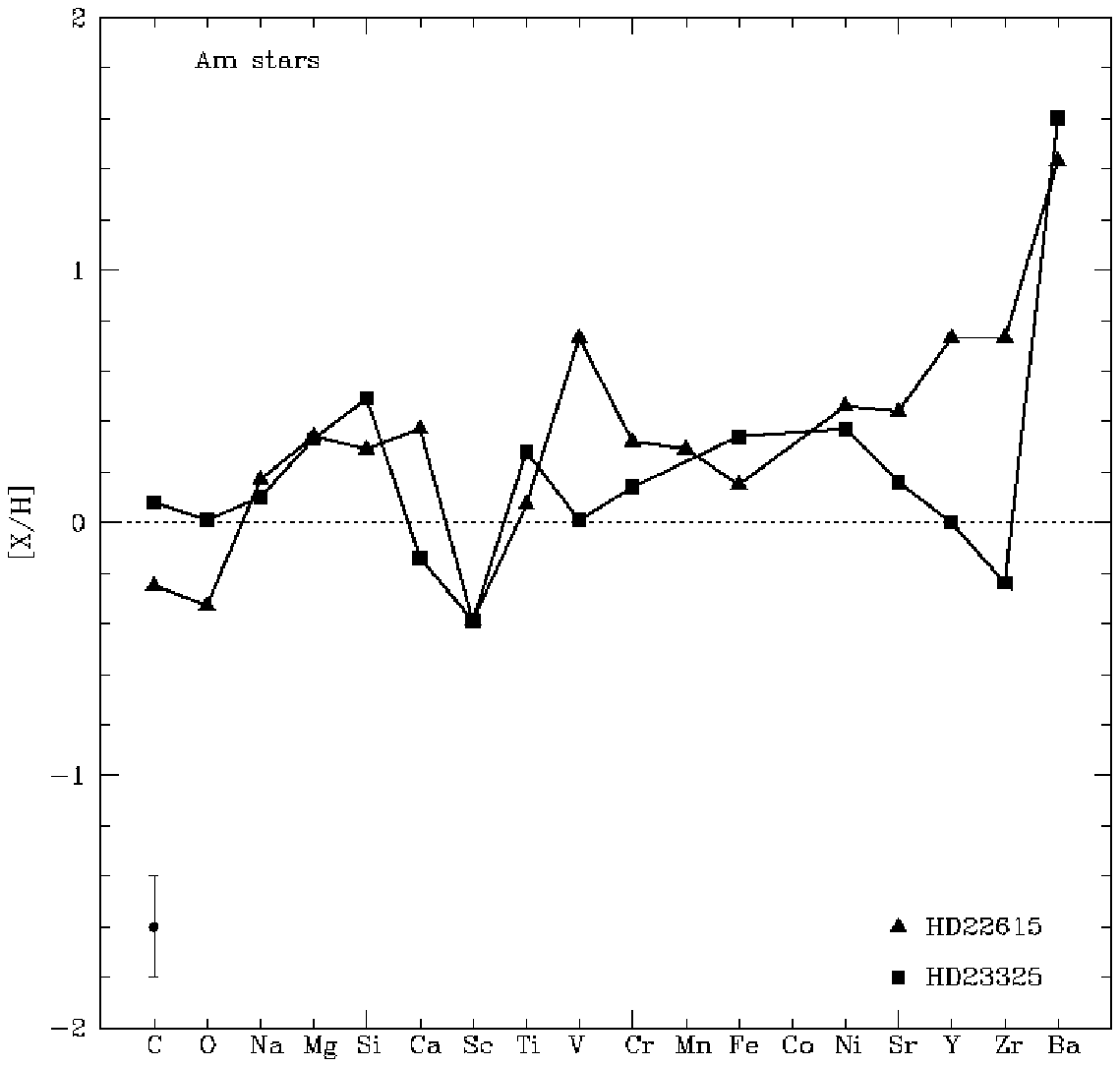}
 \includegraphics[scale=0.42]{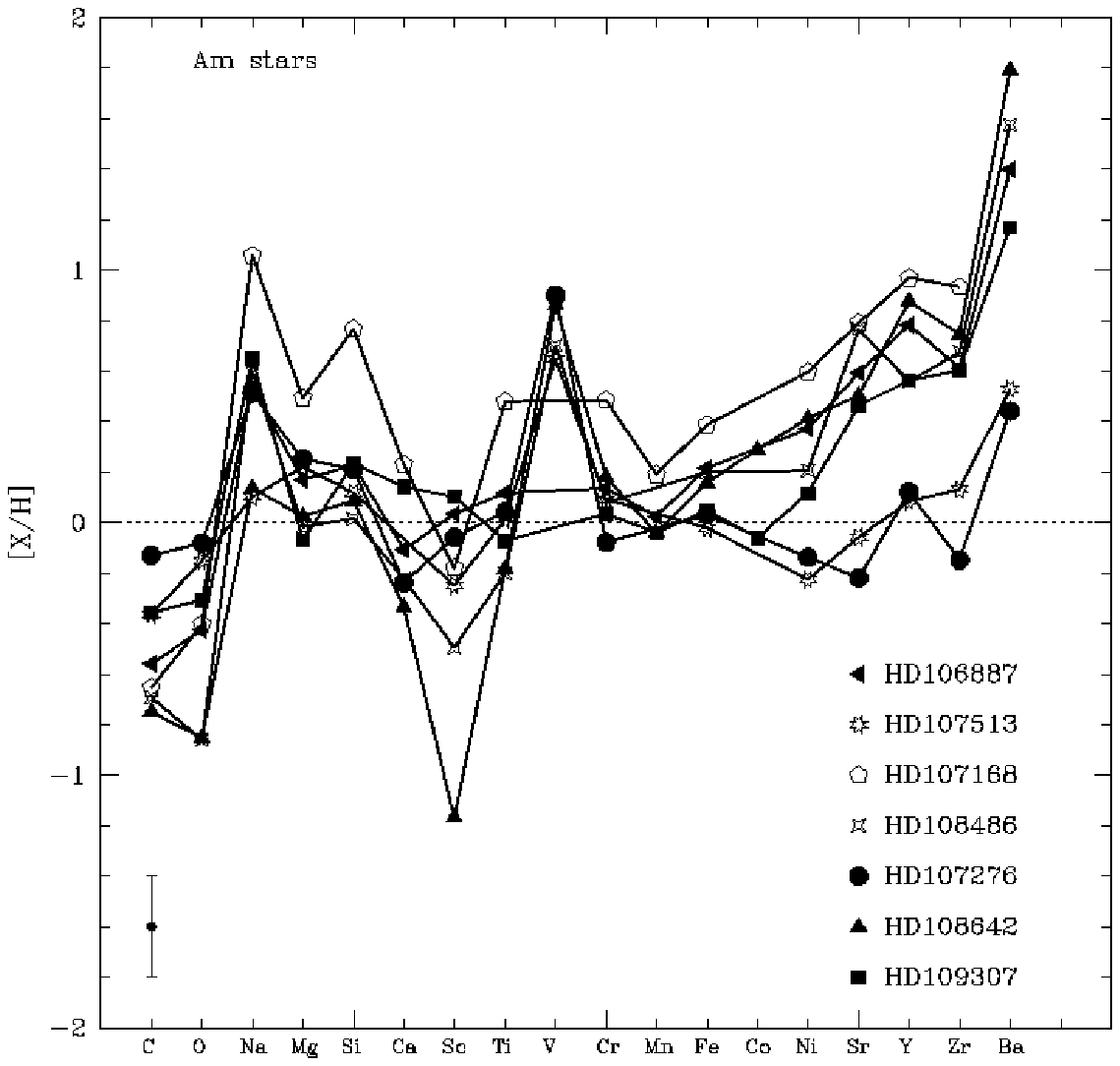}
 \includegraphics[scale=0.42]{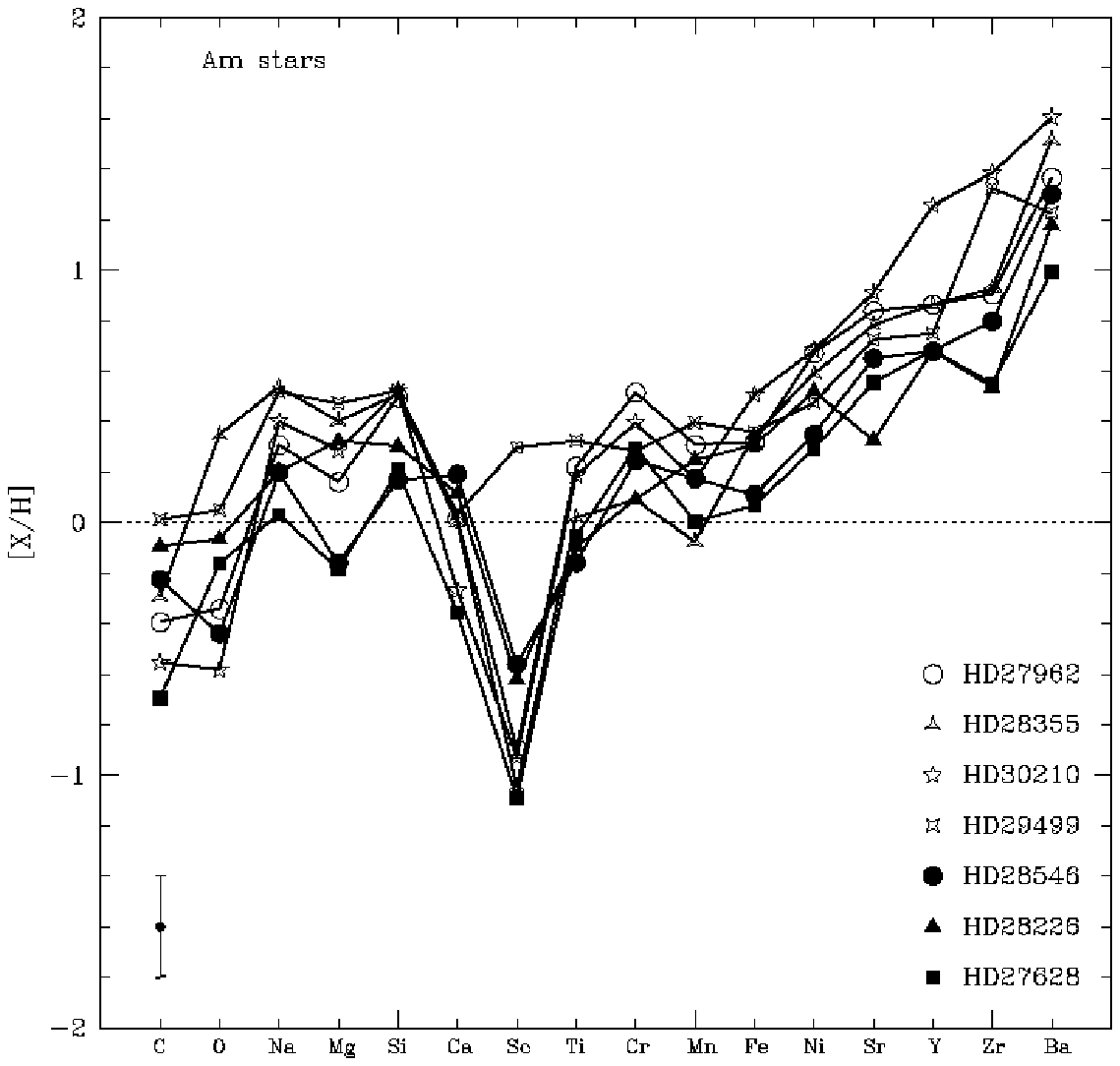}
 \caption{Abundance patterns for Am stars in the Pleiades (left pannel), Coma Berenices (middle pannel) and Hyades (right pannel) clusters.}
        \label{figure_mafig}
    \end{figure}

We have derived abundances of several chemical elements using synthetic spectra in 21 A/F stars members of the Pleiades (100 Myr), 22 A/F stars members of Coma Berenices (450 Myr) and 44 A/F stars members of Hyades (620 Myr) open clusters. We display in Fig.~\ref{figure_mafig} the abundance patterns for Am stars in these clusters. The typical underabundances of scandium and/or calcium and/or the overabundances of iron peak elements conspicuously show in these patterns.  \\
 We have found large star-to-star variations in abundances for several chemical elements among A stars (in contrast with the F stars). The largest spreads occur for Sc, Sr, Y and Zr while the lowest are for Mg, Si and Cr. The abundances of Ti, Cr, Ni, Sr, Y and Zr are correlated with the iron abundance.
The ratios [C/Fe] and [O/Fe] are anticorrelated with [Fe/H]. Compared to normal A stars, Am stars appear to be more deficient in C and O.  
No correlations exist between the abundances and $T_{\rm{eff}}$ nor between the abundances and $v_{e}\sin i$.\\% These behaviors were expected since the timescales of diffusion are much shorter than those of rotational mixing. The microturbulent velocity reach its maximum around mid A-type stars (2-4 km/s). Once we reach F-type stars $\xi_{t}$ decreases to the value of 1-2 km/s.\\
%Abundance determinations of A and F dwarfs in open clusters and moving groups of known properties aim at elucidating the mechanisms of mixing at play in the interiors of these main-sequence stars. The derived abundances for these stars help us to set constraints on self-consistent evolutionary models of these objects including various particle transport processes. Indeed stars in open clusters originate from the same interstellar material (i.e. they have the same age and the same initial chemical composition) and as such are very useful to test the predictions of evolutionary models.\\
The derived abundances have been compared to the predictions of recent evolutionary models. These models are calculated with the Montr\'eal stellar evolution code. Transport of chemical species includes several processes calculated from principles: radiative accelerations, turbulent diffusion, thermal diffusion and gravitational settling (for more details see Richard et al. 2001).  None of the calculated patterns reproduces fully the shape of the observed patterns. The discrepancies between derived and predicted abundances could partly be due to non-LTE effects. However, the inclusion of other hydrodynamical processes acting within the radiative zone of these stars (mixing processes/mass loss) could hinder the effects of microscopic diffusion and improve the agreement between the observations and the predictions.

\section{Gaia's predictions}
Gaia's RVS and Astrometric Field will provide us valuable information concerning the statistics of A stars in the Galaxy. Am stars represent $\sim$12\% of A stars. Simulations of RVS spectra show that we can disentangle between a normal A star and an Am star up to magnitude $G_{RVS} \sim$12-13 mag. This is due to the difference between the intensities of the calcium line in normal A and Am stars. At these magnitudes, we will have medium resolution spectra (R$\sim$11500) for more than 1 million A stars. Among these stars, a group of Am stars (the one with low calcium abundances) can be identified. Once these stars are identified, on-ground observations will be needed to acquire high resolution and high signal-to-noise spectra in order to have detailed elemental abundance analyses. On the other hand, using the astrometry data, the distances to these stars and especially for members of open clusters will be determined with better accuracies. Then, using isochrones, we will have new estimation about the ages of the clusters and more constraints for the evolutionary models. For a magnitude V$\sim$12-13 mag and for a typical A star, we can reach out to a distance of $\sim$3 kpc. There are about 900 open clusters with d$<$3 kpc according to the WEBDA database.

%\begin{acknowledgements}

%\end{acknowledgements}

%%-----------------------------
%%   Bibliography
%%-----------------------------
%
% The reference list should contain all the references cited in the text, ordered alphabetically by surname (with
% initials following). If there are several references to the same first author, they should be entered according
% to the following scheme:
% 1. One author: chronologically
% 2. Author, one co-author: alphabetically by co-author, then chronologically
% 3. Author, two or more co-authors: chronologically.
%
% Please note that for papers that have more than five authors, only the first three should be given, followed
% by "et al."
%
% The format for references is the one adopted by A&A (see the example below).
%
% To set the reference list in the proper A&A format, we encourage you to use BibTEX and the natbib
% package instead of the standard thebibliography environment.
%

%
\end{document}